\documentclass[preprint,12pt]{elsarticle}

\usepackage{graphicx}
\usepackage{hyperref}
\usepackage{amssymb}
\usepackage[ruled]{algorithm2e}

\newcommand{\sgn}{\mathop{\mathrm{sgn}}}

\def \beq {\begin{equation}}
\def \eeq {\end{equation}}

\begin{document}

\begin{frontmatter}



\title{Chaotic oscillations in singularly perturbed
FitzHugh-Nagumo systems}


\author{Peterson T.C. Barbosa}
\ead{petersontaylor@gmail.com}
\author{Alberto Saa}
\ead{asaa@ime.unicamp.br}
\address{Departamento de Matem\'atica Aplicada,\\
Universidade Estadual de Campinas,
13083-859 Campinas,  SP, Brazil}

\begin{abstract}
We consider the  singularly perturbed limit of periodically excited  two-dimen\-sional
FitzHugh-Nagumo systems. We show that the dynamics of such systems are essentially 
 governed  by an one-dimensional map and present a numerical scheme to accurately compute  it  together with its Lyapunov exponent. We then investigate the occurrence of chaos by varying the parameters of the system, with especial 
 emphasis on the simplest possible chaotic oscillations. Our 
 results corroborate and complement some 
 recent works on bifurcations and routes to chaos in 
 certain particular cases corresponding to piecewise linear FitzHugh-Nagumo-like systems. 
\end{abstract}


\end{frontmatter}


\section{Introduction}
The FitzHugh-Nagumo system is a simplified model for 
 nerve conduction which has become
a paradigm for dynamical and neuronal investigations since,
despite of being considerably simple, it is still able to reproduce qualitatively   the main
dynamical behaviors of realistic neural models.
 The literature on this
subject is quite vast and, in many times, rather confusing with respect to definitions and names. 
We adopt here the nomenclature and the main conventions of \cite{Izhi}. 
The FitzHugh-Nagumo model for a single externally excited neuron is given by the
following 
two dimensional non-autonomous system of first order differential equations 
\beq
\label{fhn}
\left\{ 
\begin{array}{l}
\varepsilon \dot v = f(v) - w + \psi(t) \\
\dot w = v - \delta w
\end{array}
\right.
\eeq
where $\delta$ and $\varepsilon$ are non-negative
parameters and $\psi(t)$ stands for the external excitation. The usual and simplest choice for the function $f(v)$ is
\beq
f(v) = v -\frac{v^3}{3},
\eeq
and we also adopt it hereafter. Our approach, however, can be easily modified to deal with any other kind of function $f(v)$. Typically, the $\varepsilon$ parameter in the FitzHugh-Nagumo system is considered to be small, implying the usual dynamical decomposition in fast $(v)$ and slow $(w)$ modes 
 \cite{Izhi}.  In the absence of the external excitation ($\psi(t)=0$), the intersection
 of the nullclines $w=f(v)$ and $v=\delta w$ determines the possible fixed points for the system. Since we are mainly interested in the simplest possible chaotic solutions for the system (\ref{fhn}), we will restrict ourselves to the cases exhibiting only one fixed point, which corresponds to the values of
 $  \delta <1 $. The three-dimensional  phase space of the non-linear system (\ref{fhn})    is known to  accommodate a rich dynamical behavior, ranging from periodical bursting to chaotic oscillations. 

The limit  $\varepsilon\to 0$ is called a singular perturbation  of (\ref{fhn}) since we are, effectively, reducing the phase space dimension of the system (see, for instance, \cite{OthmerXie1,PRE}). In such a limit, which preserves many dynamical properties of the original system, the variable $v$ is constrained to move along the nullcline $\dot{v}=0$, implying that the dynamics of the system are such that
\beq
\label{constraint}
w(t)=f(v(t)) + \psi(t),
\eeq
for all $t$.
By substituting (\ref{constraint}) in the second equation of (\ref{fhn}), we have
\beq
\label{eq}
\frac{d}{dt}\left( f(v) + \psi(t)\right)  = v-\delta \left( f(v) + \psi(t)\right), 
\eeq
which is a first order non-autonomous differential equation with, consequently, 
a two-dimensional phase space.  Equation (\ref{eq}) is not separable for 
generic $\psi(t)$ and, hence, there is no hope of obtaining its general solutions.  Nevertheless,  
since it  effectively corresponds to a two-dimensional system, one might ask at this point if the phase space of the singularly perturbed FitzHugh-Nagumo equation (\ref{eq}) would not be classified according
to the Poincar\'e-Bendixson theorem, what would leave no room for any chaotic dynamical behavior.  The answer is definitely no. Equation (\ref{eq}) can be written as
\beq
\label{eq2}
(1-v^2) \dot{v} = (1-\delta)v- \dot{\psi}(t)+\delta \left(\frac{v^3}{3} - \psi(t)\right), 
\eeq
from where we see clearly that the system is singular on the vertical (on the plane $(v,w)$) lines $v=\pm 1$. Thus, we cannot use the Poincar\'e-Bendixson theorem, nor its recent generalization 
for piecewise Lipschitz continuous systems \cite{Melin},   globally on the phase space of (\ref{eq2}). As we will see, Equation
(\ref{eq2}) does indeed exhibit chaotic behavior, even for its simplest forms.

There is a well known and studied particular case for which we do have analytical solutions for (\ref{eq}). It corresponds to a neuron excited by periodic instantaneous (rectangular) pulses of the form
\beq
\label{pulse}
\psi(t) = \left\{
\begin{array}{ll}
0, & {\rm\ for\ } 0\le t < \theta, \\ 
A, & {\rm\ for\ } \theta \le t < T,
\end{array}
\right.
\eeq
with $\psi(t\pm mT) = \psi(t)$, $m=1,2,3\dots$, and non-negative $A$. This kind of external
excitation is physiologically reasonable and has been considered before in numerous analysis of the FitzHugh-Nagumo system, see, for instance, \cite{OthmerXie1,PRE,NonlDyn} and the references cited therein. In particular, we  will follow here the approach of \cite{NonlDyn}  and introduce the one-dimensional map $v_{n+1} = F(v_n)$ corresponding to a stroboscopic map\cite{Izhi} of the equation (\ref{eq}) sampled on periods $T$, {\em i.e.}, if $v_0 = v(0)$, then $v_n = v(nT) = F^n(v_0)$. This kind of map can be used to explore many aspects of the underlying dynamics of the system.  
For a comprehensive recent review on maps in neuronal dynamics, see \cite{Maps}. 
In the next section, we show how to determine numerically and accurately the map $F(v)$ together  its Lyapunov exponent, which will be used eventually to investigate  the
  main point of the present work: the chaotic oscillations of (\ref{eq}) under an external excitation of the type (\ref{pulse}).

\section{The FitzHugh-Nagumo map and its Lyapunov exponent}

The stroboscopic map $F(v)$ for the equation (\ref{eq}) requires the evaluation of the solutions  $v(t)$ 
at $t=T$, for generic values of $v_0=v(0)$. We do not
adopt here any further simplification as, for instance, those ones considered in
 \cite{OthmerXie1,PRE,NonlDyn} corresponding to assume a piecewise linear ``N"-shaped $f(v)$ function, which makes equation (\ref{eq}) linear and, consequently, simplifies
 considerably the problem.
Far from the discontinuities of $\psi(t)$, equation (\ref{eq2}) reads
 \beq
 \label{eq1}
 (1-v^2)\dot{v} = \frac{\delta}{3} v^3 + (1-\delta)v - \delta A,
 \eeq
  for $\theta < t < T$,   whereas for $0< t <\theta$ we have exactly  the 
  same expression, but with 
  $A=0$. Equation (\ref{eq1}) can
   be easily integrated by a simple quadrature, leading to the following general solution
 \beq
 \label{int}
 \int_{v_0}^v \frac{1-s^2}{\frac{\delta}{3} s^3 + (1-\delta)s - \delta A}\, ds = t-t_0.
 \eeq
In order to obtain the matching conditions for $v(t)$ on the discontinuities  of $\psi(t)$, let us integrate (\ref{eq}) in the distributional sense, for instance, from $t=\theta - \epsilon$ to $t=\theta + \epsilon$, and then take the limit $\epsilon \to 0$. One gets
\beq
\label{jump1}
f(v(\theta^+)) = f(v(\theta^-)) - A.
\eeq
Analogously, we will have
\beq
\label{jump2}
f(v(0^+)) = f(v(0^-)) + A.
\eeq
The dynamical picture in the phase space  of (\ref{eq}) is as follows. Far from the singularities of
$\psi(t)$, the solution is constrained to move on the surface (\ref{constraint}). When a
pulse appears (for instance, for $t=\theta$), the solution jumps instantaneously (same $t$) and horizontally (same $w$) to a new value of $v$ according to the cubic equation (\ref{jump1}), which in fact corresponds to jump horizontally from the curve   on the plane $(v,w)$ given by (\ref{constraint}) with
$\psi(t)=0$ to the new one with $\psi(t)=A$. This situation is well illustrated, for the case of   piecewise linear ``N"-shaped $f(v)$ functions, in  \cite{OthmerXie1} and \cite{NonlDyn}. The present situation is completely analogous to those ones.

There are others discontinuities in the dynamics of $v$. As it was already noticed, an inspection of (\ref{eq1})
reveals that $\dot{v}$ diverges for $v=\pm 1$. Assuming a $\delta$ parameter small enough to assure that the right handed side of (\ref{eq1}) has, for a given value of $A$, only one zero and that it is located in the interval
$(-1,1)$, we will have $\dot{v}<0$ for $v>1$ (respectively, $\dot{v}>0$ for $v<-1$).   A solution starting, say, at $v>1$ is driven towards the point $v=1$. On this point, since $\dot{v}$ diverges on it, the solution jumps instantaneously and horizontally to the point $v=-2$, and then it is driven towards $v=-1$ and so on, giving origin to a periodic oscillation in the well known fashion. Notice that, inside the interval $(-1,1)$, the directions are reversed, {\em i.e.}, 
$\dot{v}>0$ for   $v>v_*$ (respectively,  $\dot{v}<0$ for  $v<v_*$), where $v_*$ stands for the fixed point of (\ref{eq1}), which is unique by hypothesis. In this way, a solution starting at $v>v_*$ will be
driven towards $v=1$ and start a new periodic solution, jumping to $v=-2$ and so on.
Solutions starting in the regions $|v|>2$ will be driven towards the singularity
$v=\sgn(v)$ and also start the periodic oscillations.
One can, without loss of generality, remove the regions  $-1<v<1$ and $|v|>2$ from our map, since no oscillating  solution is ever injected there.

For generic $\delta < 1$ and $A>0$, the integral (\ref{int}) can be evaluated by expanding the rational fraction in term of partial fractions. (See the Appendix A for these details, for the numerical scheme, and for the algorithm.) However, for $\delta\to 0$ (the van der Pol limit of the FitzHugh-Nagumo equation), the problem is greatly simplified, despite of keeping unaltered its main dynamical properties. In such a limit, we have simply 
\beq
\label{vdp}
\ln \frac{v}{v_0} - \frac{v^2-v_0^2}{2}  = t - t_0 .
\eeq
In fact, the main dynamical properties of the model are independent of the details of the function $f(v)$, provided it has the usual and characteristic ``N" shape. The solution
for a FitzHugh-Nagumo equation with a generic $f(v)$ under excitations of the type (\ref{pulse}) is
given by
 \beq
 \label{int1}
 \int_{v_0}^v \frac{f'(s)}{ s-\delta( f(s) + A)}\, ds = t-t_0.
 \eeq
The matching and the jumping conditions for $v(t)$ are essentially the same. Our stroboscopic map can be easily modified to incorporate the general quadrature (\ref{int1}), which might be, if necessary, evaluated numerically with good accuracy.

We have set up a numerical code \cite{scilab}, using Scilab, that takes into account the jumps $v=\pm 1 \to v=\mp 2$ and the match conditions (\ref{jump1}) and (\ref{jump2}) in
the general solutions (\ref{int}). The pertinent  details are presented in the Appendix A. We have used it to construct accurately the stroboscopic map 
$v_{n+1}=F(v_n)$ for any values of the parameters $\delta$, $A$, $\theta$, and $T$.
\begin{figure}[htb!]
\centering{
\includegraphics[width=0.740\linewidth]{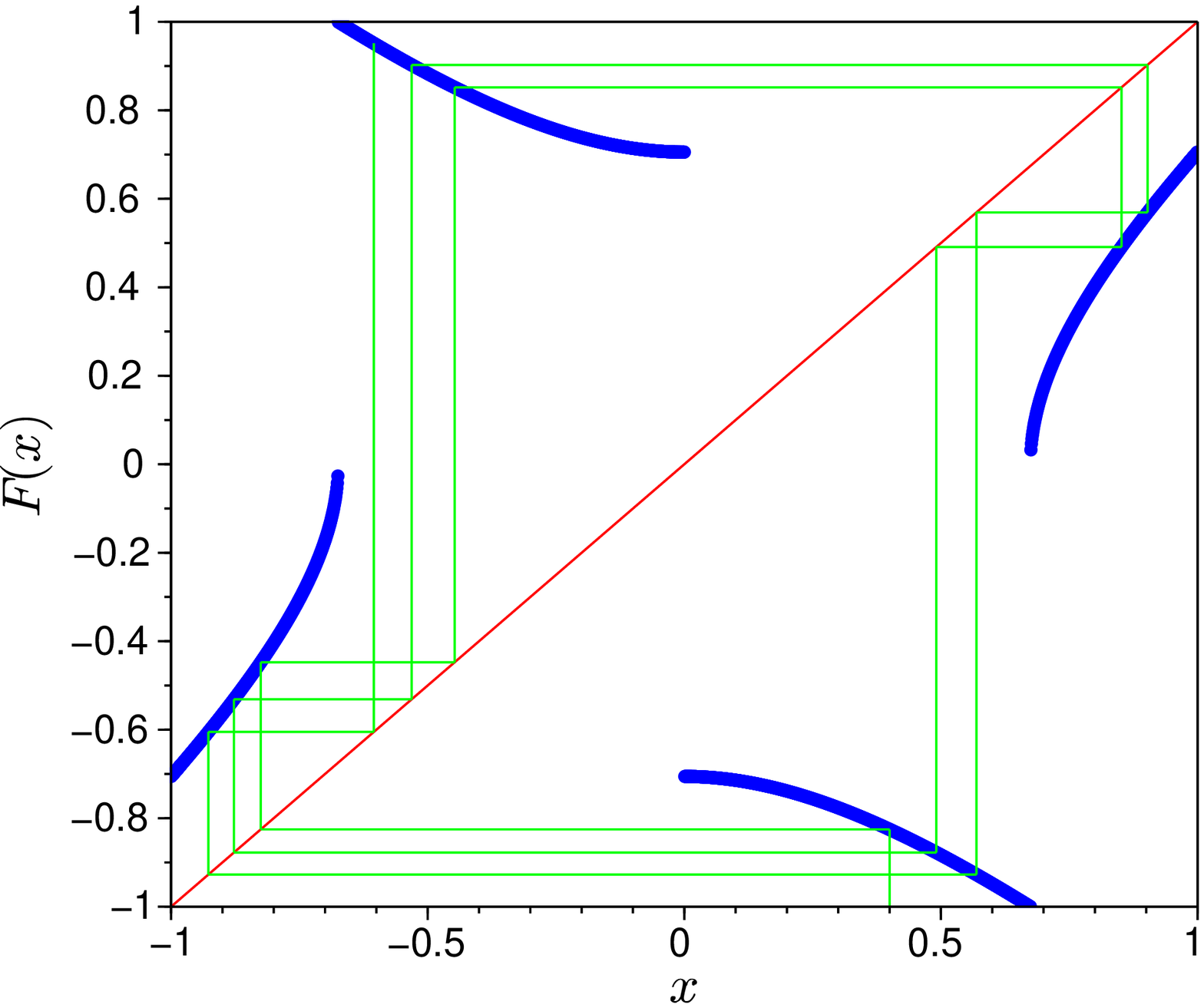}\\
\includegraphics[width=0.740\linewidth]{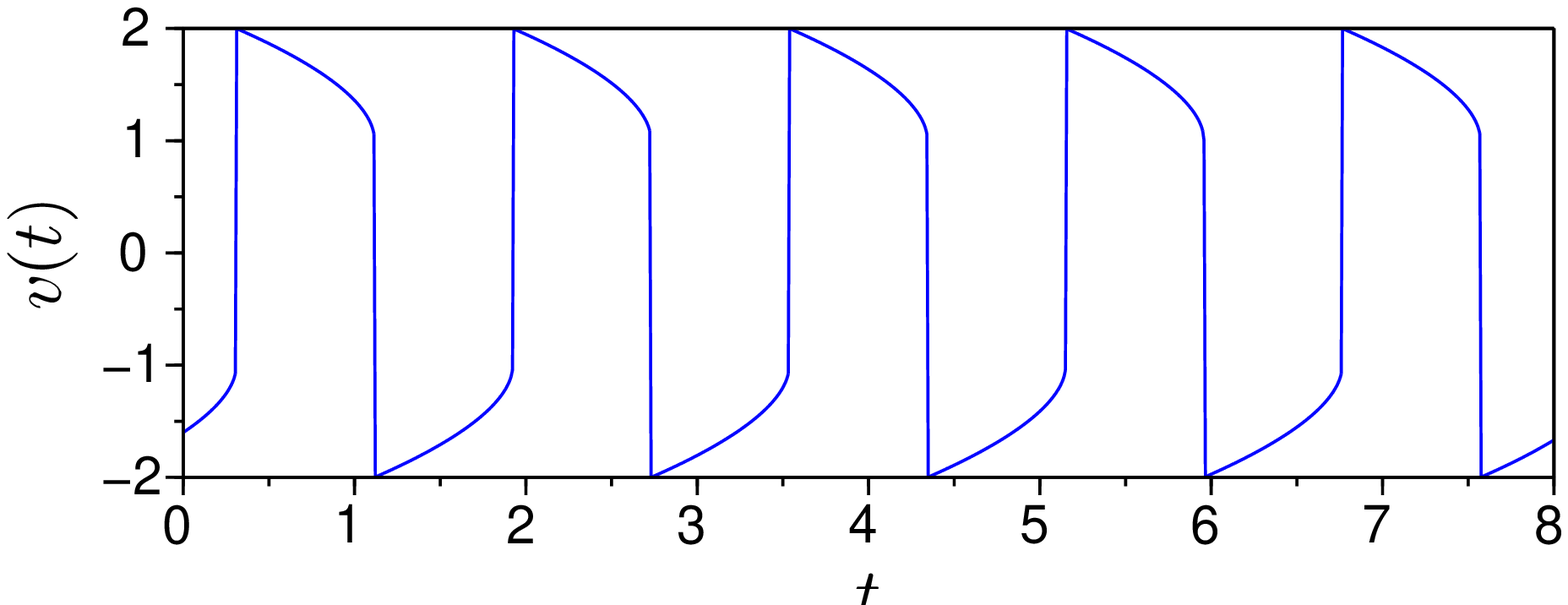}
}
\caption{The stroboscopic map for $\delta=A=0$ and $T=2$ (top), with 11 cobweb iterations represented,  and the 
corresponding periodic solution of (\ref{eq}) (bottom).}
\label{fig1}
\end{figure}
Figure \ref{fig1} depicts a typical map for $\delta = A=0$. This particular situation, of course, corresponds to a regular periodic solution (no external excitation). Provided the value of $\delta$ be such that there is only one fixed point and it is located in the interval $(-1,1)$, the dynamical behavior seems to be essentially independent of $\delta$. Since we have removed the region $(-1,1)$
of our analysis, we redefine our intervals in order to have 
$[-2,-1]\cup[1,2]\to [-1,1]$ by simple shifts.  The obtained maps are typically piecewise smooth and  the 
interval $[-1,1]$ can be effectively  partitioned into several dynamically disjoint
and inequivalent  sets. 

Since we are dealing with piecewise smooth one-dimensional maps $F(x)$, we could in principle study their dynamical properties from 
their Lyapunov exponent 
\begin{equation}
\lambda_x=\lim_{n\to\infty}\frac{1}{n}\sum_{k=0}^{n-1}\ln\left|F'\left(F^{k}(x)\right)\right|.
\label{lamba}
\end{equation}
 A positive Lyapunov exponent (\ref{lamba}) would correspond undoubtedly to chaotic oscillations in this context. Since our maps are piecewise smooth, it is indeed possible that the interval $I=[-1,1]$  be partitioned into regions with different Lyapunov exponent for the same map. The index $x$ in (\ref{lamba}) is a reminder of this important fact. The map depicted in Fig. \ref{fig1}, for instance, should have a vanishing Lyapunov exponent
 for all the interval $I=[-1,1]$, and
 we could verify it numerically with very good accuracy. This map, in particular, corresponds to a periodic solution with period 
$T_0 = 3 -2\ln 2$, but sampled with period $T=2$. 
Since $T/T_0$ is irrational, we do not 
expect either fixed or periodic points of any order, {\em i.e.}, there should exist no $n>0$ such that $x_*=F^n(x_*)$. Since the underlying solution is regular and periodic, 
we expect neither 
positive Lyapunov exponents for this situation. However, by varying the map parameters 
we can easily induce the apparition of fixed ($x_*=F(x_*)$) and periodic  points. If a fixed point
is stable $(|F'(x_*)|<1)$,
its basin of attraction ${\cal B} = \{x\in I | \lim_{n\to\infty} F^n(x)\to x_* \}$ will have a Lyapunov exponent $\lambda_x = \ln |F'(x_*)| < 0.$ On the other hand, 
unstable fixed points 
 $(|F'(x_*)|>1)$ are typically good candidates to induce chaotic dynamics. Neutral
 fixed points $(|F'(x_*)|=1)$ could also in principle give origin to chaotic behavior. We will return to this point in the last Section.

\begin{figure}[tb]
\centering{
\includegraphics[width=0.740\linewidth]{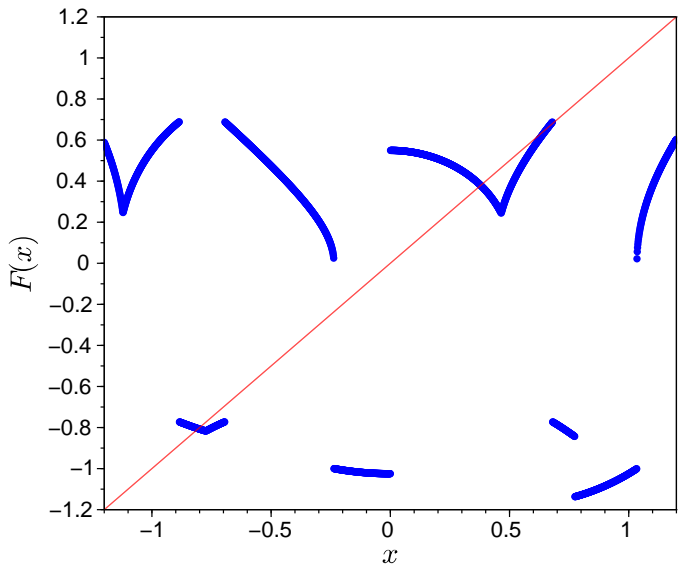}
}
\caption{The stroboscopic map for $\delta=0$, $A=3/4$, $\theta=1/2$, and $T=4$. Notice the presence of three fixed points $x_*=F(x_*)$. Fig. \ref{fig3} depicts these fixed
points separately with some cobweb iterations.}
\label{fig2}
\end{figure}

\begin{figure}[tb]
\centering{
\includegraphics[width=0.49\linewidth]{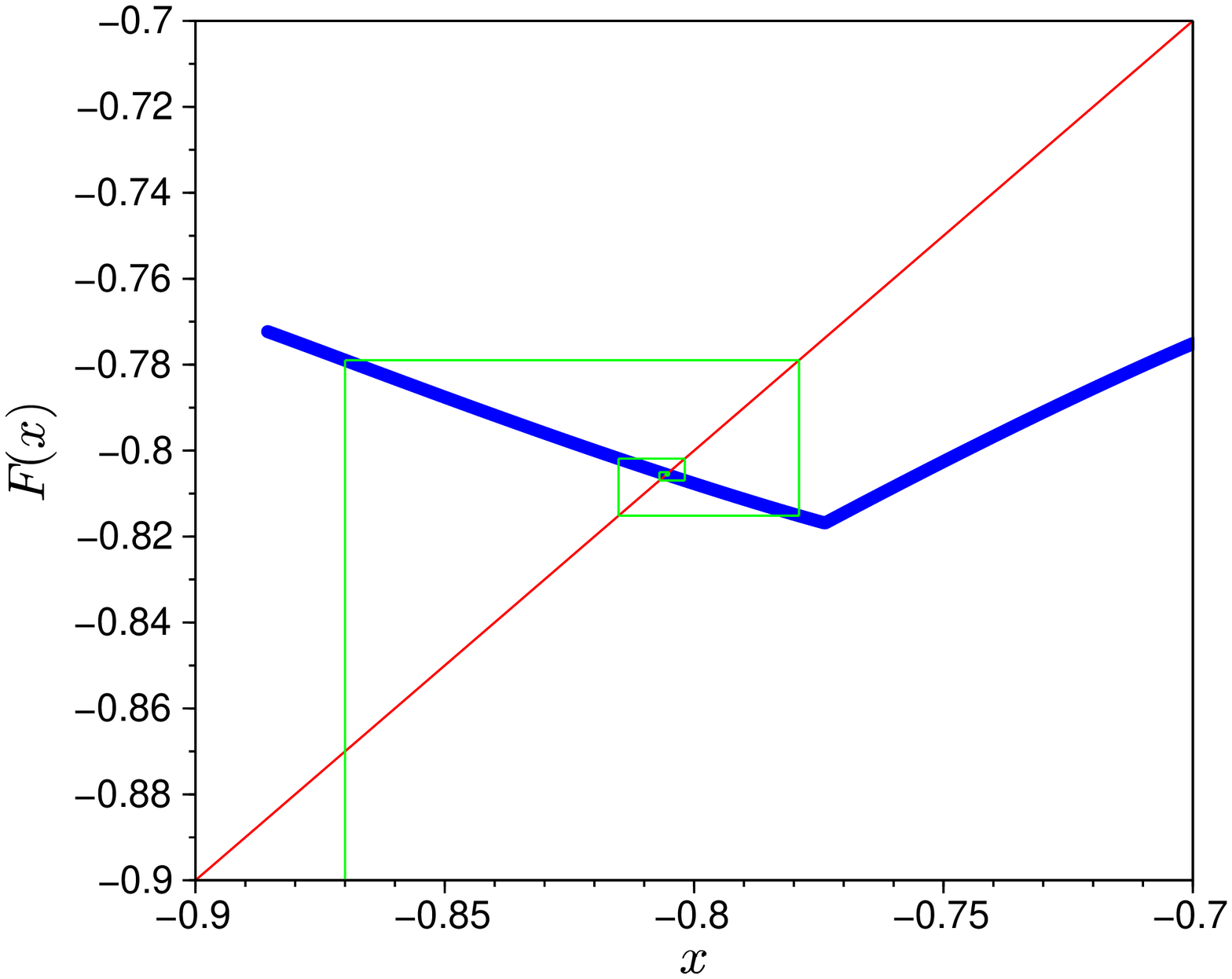}
\includegraphics[width=0.49\linewidth]{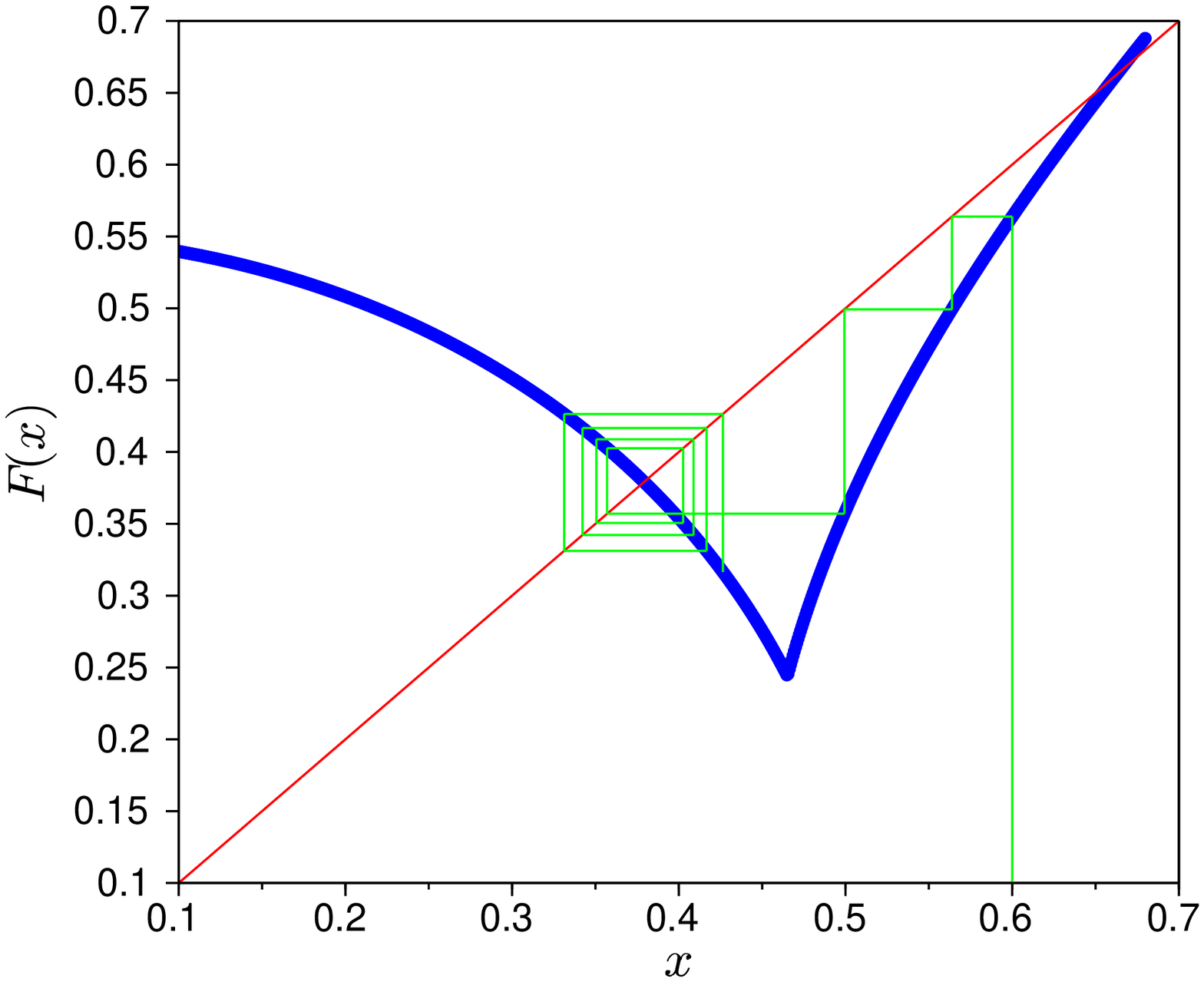}
}
\caption{The
fixed points for the  stroboscopic map of Figure \ref{fig2}
with 11 cobweb iterations represented. Left: the stable fixed point. Right: the two unstable fixed points.}
\label{fig3}
\end{figure}

Figure \ref{fig2} 
depicts the stroboscopic map for the case with $\delta=0$, $A=3/4$, $\theta=1/2$, and $T=4$, which clearly exhibits three fixed points, which are
shown separately  in Figure \ref{fig3}, with some cobweb iterations.
 The first fixed point (ordered according to the value of $x$) is clearly a stable one. For any point $x$ of its basin of attraction ${\cal B}$, we will have 
$\lambda_x = \ln |F'(x_*)| = -0.965\dots$.  The dynamical behavior is rather clear. The solutions starting in ${\cal B}$ at $t=0$, after undergoing a transient regime (the oscillations far from the fixed point $x_*$), will tend to $x_*$, {\em i.e.}, they will tend towards a periodic behavior with period $T$. The forcing term $\psi(t)$
is clearly driving the solution in this case, a rather common behavior for damped excited oscillators.  
 Much more interesting, however, are the other two fixed points. They are clearly unstable $(|F'(x_*)|>1)$ and the solutions tend to depart from each other, see the cobweb iterations depicted in Fig. \ref{fig3}. However, the map is such that the trajectory is constantly re-injected in the region with $|F(x)|>1$, leading eventually to a positive Lyapunov exponent for this region of the interval $I$. For this case in particular, one has $\lambda_x = 0.289\dots$. A map with $|F(x)|>1$  implies exponential deviations for initially nearby trajectories, leading in this way to the most distinctive  characteristic of chaotic dynamics: the extreme sensitivity to initial conditions. 
 
\begin{figure}[tb]
\centering{
\includegraphics[width=0.74\linewidth]{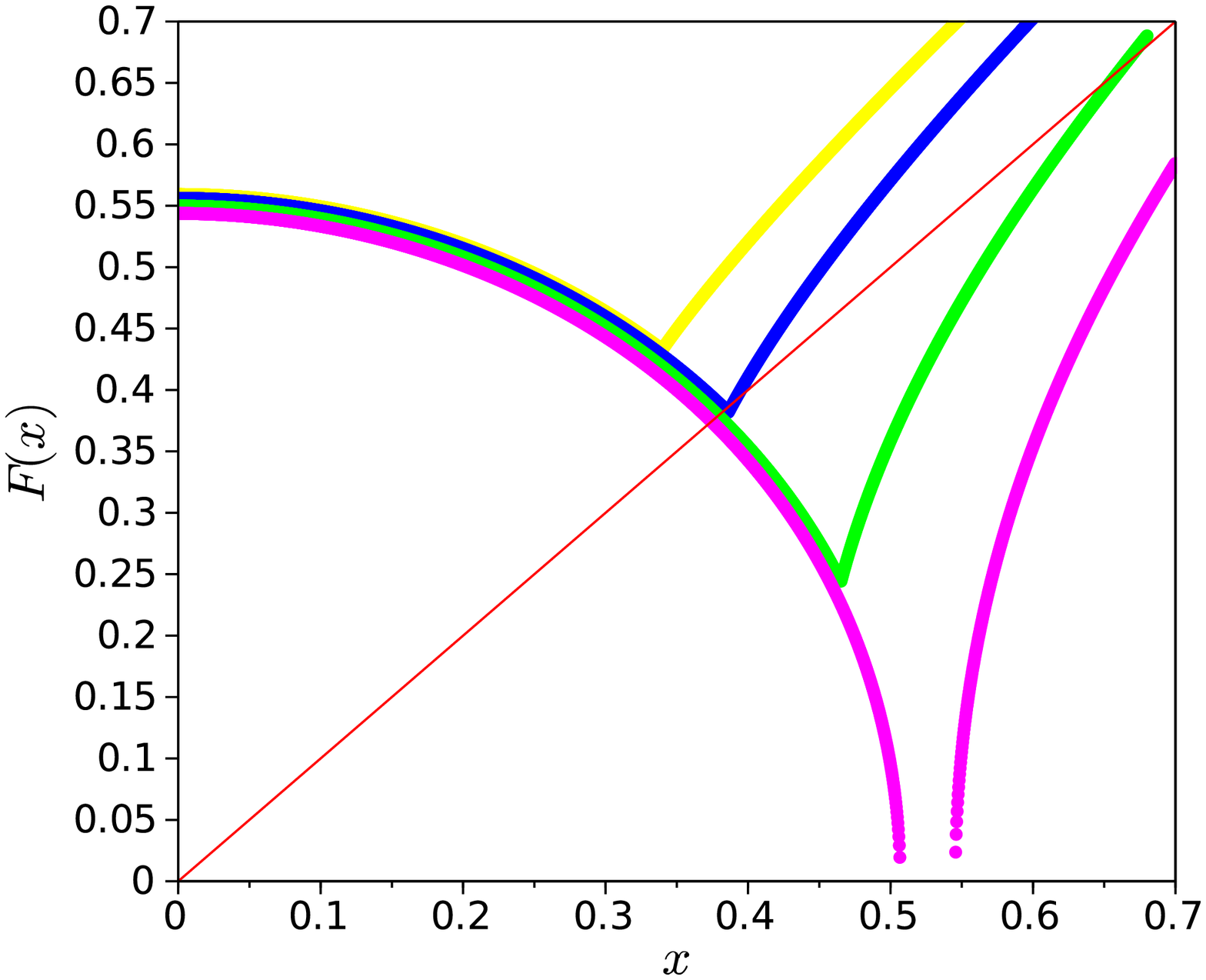}}
\caption{Detail of the unstable fixed points for the stroboscopic map with
$\delta =0$, $\theta=3/4$, and $T=4$. From top to bottom, the curves corresponds to
$A=0.6$ (no unstable points, regular motion), $0.65$ (onset of chaos), $0.75$ (two unstable points, chaos), and $0.85$ (discontinuous map, regular motion). See the text for further details. }
\label{fig4}
\end{figure}

 Our numerical set up allows also the investigation of the origin of the unstable fixed points in the stroboscopic map and, consequently, the onset of the chaotic behavior in 
 the FitzHugh-Nagumo system, as the system parameters vary. An example is shown in Figure \ref{fig4}.
We keep $\delta =0$, $\theta=3/4$, and $T=4$ fixed, and vary $A$. The onset of chaos corresponds to the appearance of the unstable fixed points, which occurs for $A=A_0\approx 0.65$, the second curve, from top to bottom, in Figure \ref{fig4}. By increasing the value of $A$, the distance between the two fixed points increases, enlarging in this way the chaotic region, which corresponds essentially to the region bounded by the two unstable fixed points. The corresponding Lyapunov exponent $\lambda_x$ also increases, indicating a stronger chaotic behavior. However, for $A=A_1\approx 0.83$, the map becomes discontinuous in this region, the branches corresponding to the two fixed points separate from each other and a gap between them appears. The trajectories that were typically chaotic now tend to a limit cycle in this gap, and the system becomes rather abruptly a regular one with a vanishing Lyapunov exponent. The last curve in  Figure \ref{fig4} illustrates namely this situation. The overall dynamical picture is the following. For $0<A<A_0$, the system is regular, {\em i.e.}, there is no chaotic motion, there is no sub-interval with positive Lyapunov exponent. For this case, typically, we have subintervals with vanishing and negative Lyapunov exponents. The former corresponds to those situations for which the external excitation is not strong enough to disturb considerably the
natural
 periodic  solutions of the FitzHugh-Nagumo system. On the other hand, the latter corresponds to the appearance of stable fixed points, indicating that the external excitation could effectively be  driving the system. For $A_0<A<A_1$, we have new dynamical behavior associated to the regions with positive Lyapunov exponents, namely the regions limited by the unstable fixed points. We still have, however, coexisting  
  regions with vanishing and negative Lyapunov exponents. For $A>A_1$, we yet have the two unstable fixed points with the solutions departing from each other, but due to the gap between them, the solutions are not anymore  continuously re-injected in regions with
$F'(x)>0$. The system ceases to exhibit chaotic motion in this case. It is also noticeable that the basin of attraction of the stable fixed points grows in size for increasing values of $A$. This is hardly a surprise since, with larger amplitudes,  the external excitation tends to take over  the dynamics of the system more easily. Such an overall behavior seems to be quite generic and robust against small variations in $\delta$, $\theta$, and $T$, and the shape of $f(v)$ as well. We have also detected a property of the system which dynamical origin 
 is still unclear for us: smaller values of $\theta$ seems to favor the appearance of unstable fixed point leading to chaos in  the FitzHugh-Nagumo system.
  \begin{figure}[tb]
\centering{
\includegraphics[width=0.74\linewidth]{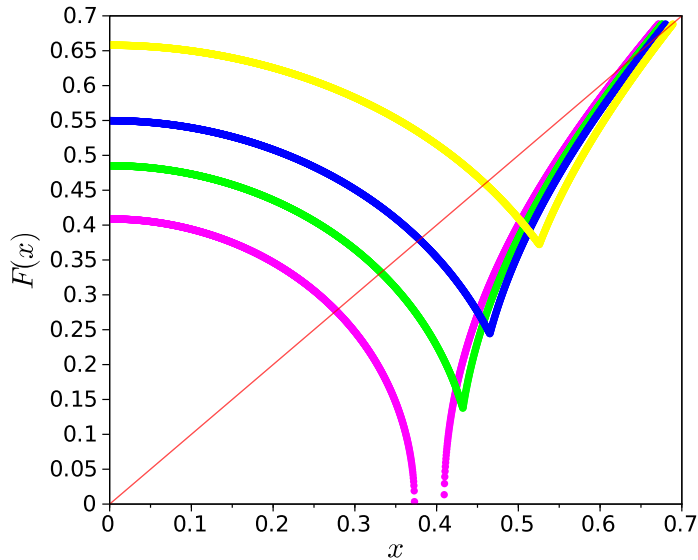}}
\caption{Detail of the unstable fixed points for the stroboscopic map with
$\delta =0$, $A=3/4$, and $T=4$. From left to right, the curves corresponds to
$\theta=0.45$ (discontinuous map, regular motion), $0.475$, $0.5$, and $0.55$ (two unstable points, chaos). The onset of chaos corresponds to $\theta\approx 0.463$. It is interesting to note that variations in $\theta$ imply deformations on the stroboscopic map mainly along the right branch, while for variations in $A$ the deformations are along the left branch of the map, compare with Fig. (\ref{fig4}).  }
\label{fig5}
\end{figure}
Moreover, the route to chaos associated with the  the parameter $\theta$ (and the period $T$) seems to be essentially the same  associated to $A$. 
Figure (\ref{fig5}) depicts the situation of the unstable fixed points for different values of $\theta$. The situation is very similar to that one depicted in (\ref{fig4}). The routes to chaos, in particular, are complementary in the sense that both routes are associated to deformations in the stroboscopic map around the unstable fixed point. Variations in $A$ imply deformation mainly along the left branch of the map (Fig. (\ref{fig4})), while variations in $\theta$ do it along the right branch (Fig. (\ref{fig5})). These results and scenarios are, again, quite robust against variations in $\delta$, provided they do not change the number of the fixed points in the original FitzHugh-Nagumo system (\ref{fhn}).

\section{Conclusion and final remarks}

\begin{figure}[tb]
\centering{
\includegraphics[width=0.49\linewidth]{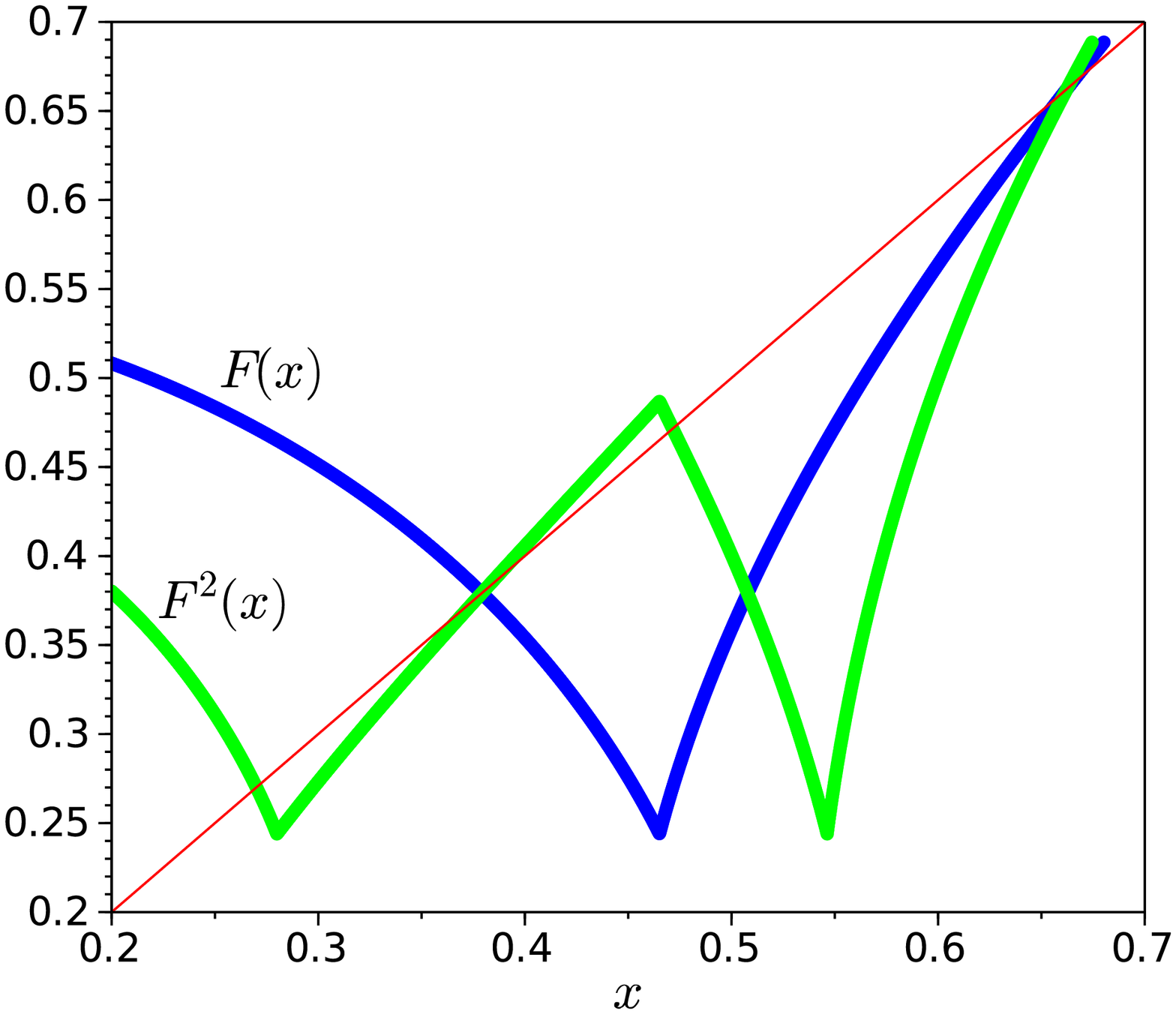}
\includegraphics[width=0.49\linewidth]{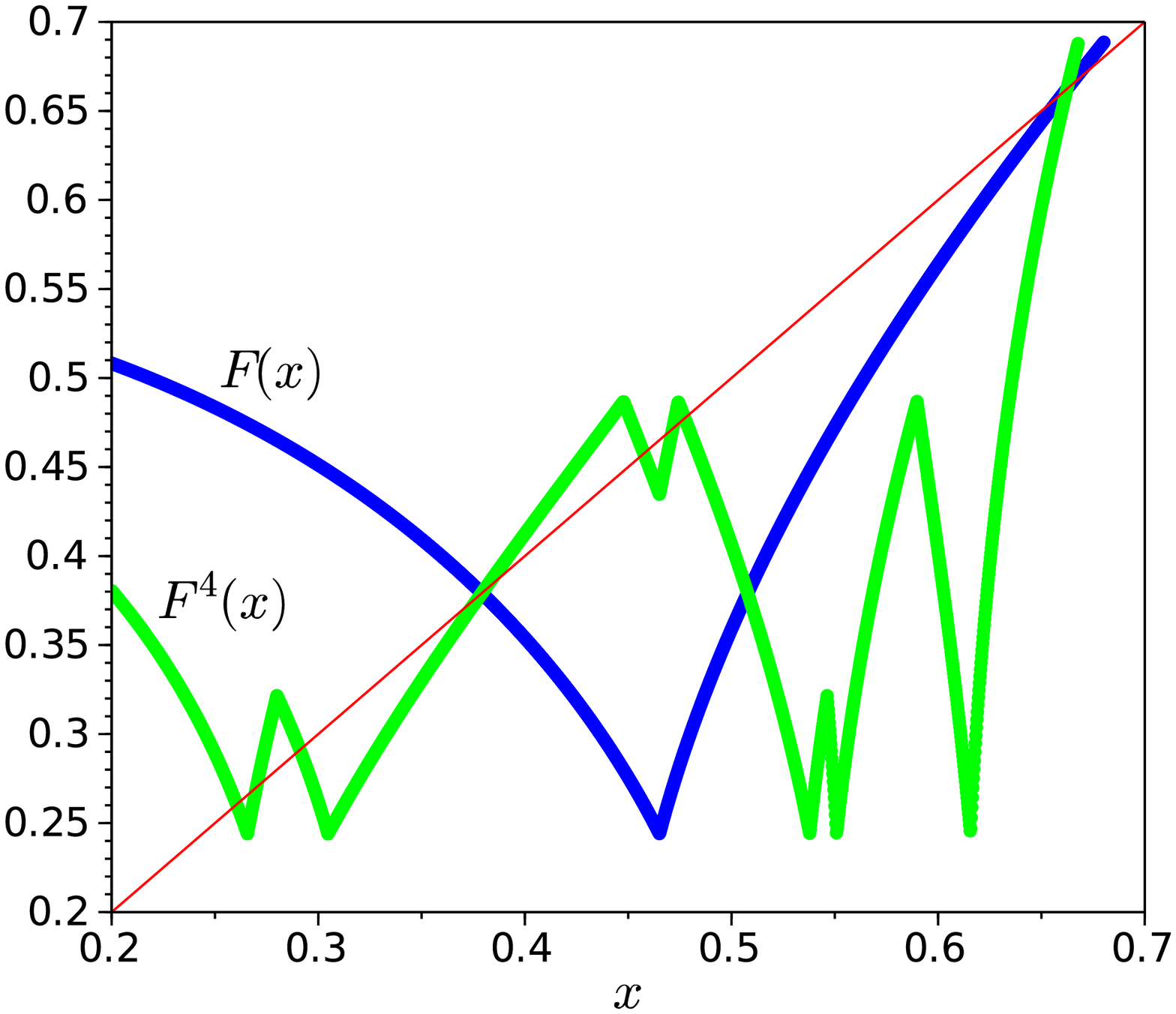}
}
\caption{Detail of the unstable fixed points for the second (left) and forth (right) iterated stroboscopic map of Figure \ref{fig2}. We note clearly the
presence of unstable periodic points, another typical feature of chaotic maps.}
\label{fig6}
\end{figure}

We have studied the singularly perturbed limit of periodically excited 
 two-dimensional FitzHugh-Nagumo systems by introducing a stroboscopic one-dimensional 
 map sampled with the period of the external rectangular pulse excitation. The pertinent numerical and computational details are presented in the Appendix A. By exploring the Lyapunov exponent of the stroboscopic map, we look for chaotic behavior in the original system. We show how to identify and characterize the chaotic regions from the analysis of unstable fixed points of the stroboscopic map. Our results corroborates and, in fact, complement those ones presented in \cite{NonlDyn}, where the same analysis is employed, but for the rather common approximation of considering a piecewise linear ``N"-shaped $f(v)$ function in the Fitz-Hugh-Nagumo equation. The main emphasis in \cite{NonlDyn} was the characterization of the chaotic dynamics from
 the bifurcation structure of the stroboscopic map. Our approach  could also be used
 for this kind of analysis.
Figure \ref{fig6},
for instance, depicts the region containing the 
unstable fixed points for the second- and forth-iterated stroboscopic map of the Figure \ref{fig2} ($\delta=0$, $A=3/4$, $\theta = 1/2$, and $T=4$). One can see clearly the
presence of unstable periodic points, another typical feature of chaotic maps. Our method can be also applied to the systems 
with different functions $f(v)$ and, in particular, to that one
considered in  \cite{NonlDyn}. In fact, since they are
linear, the general solution corresponding to (\ref{int}) is trivial, but a complex dynamics do indeed arise due to the jumping and matching conditions. Our algorithm presented in the Appendix can be easily adapted for this situation, allowing the analysis of the Lyapunov exponents of the cases considered in \cite{NonlDyn}, for instance.

We finish by remarking that it is indeed possible to have chaotic dynamics for maps with neutral $(|F(x_*)|=1)$ fixed points. This is the case, for instance, of the so-called intermittent systems. The usual Lyapunov exponent vanishes for these systems, but they can still exhibit sub-exponential instabilities characterized by a positive ``subexponential Lyapunov exponent"
\begin{equation}
\lambda_x^{(\alpha)}=\lim_{n\to\infty}\frac{1}{n^{\alpha}}\sum_{k=0}^{n-1}\ln\left|F'\left(F^{k}(x)\right)\right|,
\label{lamba1}
\end{equation}
with $0<\alpha<1$, 
see \cite{SPV,SV} for further details. The parameter $\alpha$ depends on the details of the map in the vicinity of the neutral fixed point. We have identified some stroboscopic maps with fixed points which might be neutral. The possibility that these neutral fixed points could generate sub-exponential instabilities with positive $\lambda_x^{(\alpha)}$  for the FitzHugh-Nagumo system is now under investigation.

\section*{Acknowledgements}
The authors are grateful to FAPESP and CNPq for the financial support. AS wishes to thank Professor Leon Brenig for the warm hospitality at the Universit\'e Libre de Bruxelles, where part of this work was carried on.

\appendix 

\section{The numerical scheme and the algorithm}

The stroboscopic map and its Lyapunov exponent are constructed numerically from the
general  
solution (\ref{int}) and the matching condition for $v(t)$ on $v=\pm 1$ and on the discontinuities of $\psi(t)$. For $\delta =0$, the general solution (\ref{int}) is given explicitly  by the simple expression (\ref{vdp}). In order to write the solution for the general case $ 0<\delta <1$, let us consider the cubic polynomial  
\beq
\label{poly}
p_3(v) = v^3 + 3\beta v - 3A,
\eeq
with $\beta = (1-\delta)/\delta > 1$. Since we have $p'_3(v) = 3v^2+3\beta>0$, the polynomial (\ref{poly}) has only one real root. It is already in the so-called depressed form, so it is rather simple to obtain its real root $v_*$ (the fixed point) by elementary algebra.   By using the standard Vieta's substitution, one has
\beq
\label{A2}
v_* = w_* - \frac{\beta}{w_*},
\eeq
with
\beq
w_*^3 = \frac{3}{2}A + \sqrt{\frac{9A^2}{4} +\beta^3}.
\eeq
Notice that for $A=0$, we have $w_*=\sqrt{\beta}$, leading to $v_*=0$. 
We remind our hypothesis that $\delta$  must be small enough (and hence $\beta$ must be large enough) to assure that $-1< v_* <1$ for a given value of $A$.
 The polynomial
(\ref{poly}) can be factorized as $p_3(s)=(v-v_*)p_2(v)$, where the quadratic polynomial
\beq
p_2(v) = v^2 + v_*v + 3\beta+v_*^2,
\eeq
has no real solution, by construction. Finally, the rational function in the integrand
of (\ref{int}) can be expanded as 
\beq
\label{frac}
\frac{1-v^2}{\frac{\delta}{3} v^3 + (1-\delta)v - \delta A} = 
\frac{3}{\delta}\left( 
\frac{a_1}{v-v_*} + \frac{a_2v+a_3}{p_2(v)}
\right),
\eeq
where $a_1$, $a_2$, and $a_3$ are the unique solution of the linear system
\beq
\label{A3}
\left( 
\begin{array}{ccc}
1 & 1 & 0\\
v_* & -v_* & 1 \\
3\beta+v_*^2 & 0 & -v_*
\end{array}
\right)
\left(
\begin{array}{c}
a_1 \\
a_2 \\
a_3
\end{array}
\right) =
\left(
\begin{array}{c}
-1 \\
0 \\
1
\end{array}
\right).
\eeq
The case $A=0$ $(v_*=0)$ corresponds to $a_1=(3\beta)^{-1}$, $a_2=-(a_1+1) $, and $a_3=0$. By using (\ref{frac}), the integral in (\ref{int}) can be evaluated with elementary methods, leading finally to 
\beq
H(v) - H(v_0) = t-t_0,
\eeq
where
\beq
H(v) = \frac{3}{\delta}\left[
a_1\ln |v-v_*| + \frac{a_2}{2}\ln p_2(v) + 
\frac{2a_3 - a_2v_*}{\sqrt{12\beta+3v_*^2}}\arctan 
\left(
\frac{2v+v_*v}{\sqrt{12\beta+3v_*^2}}
\right)
\right].
\eeq
(Notice that the case corresponding to $\delta=0$ is given by the simpler expression (\ref{vdp}).) With the explicit function $H(v)$, one can determine the stroboscopic map by following the description of the dynamics of Section 2. This procedure is summarized in the Algorithm 1.

The Lyapunov exponent $\lambda_x$ can be evaluated from its definition for one-dimensional maps (\ref{lamba}). The derivative $F'(x)$ can be easily evaluated numerically. There is, however, a potential problem in the numerical evaluation of  $F'(x)$. Since our map is piecewise smooth and we do not know a priori the location of its discontinuities, the left and right approximations for the derivative do not always converge to the same value. In fact, we identify the discontinuities by monitoring the sudden variations in one of them.
If the left and right approximation for the derivative are close, we take the simple average as the approximation. If they differ by orders of magnitude (indeed, a factor of $2$ was enough in our computations), we  discarding the largest (in modulus) and keep the remaining one as the approximation for the derivative.

\begin{algorithm}
\SetAlgoLined
 \LinesNumbered
\KwData{$\delta$, $A$, $\theta$, $T$, $v_0$}
\KwResult{$F(v_0)$ }
$t_0 \leftarrow 0$ \;
evaluate $v_*$, $a_1$, $a_2$, and $a_3$ for $A=0$\tcp*{Equations (\ref{A2}) and (\ref{A3})}
$t_1 \leftarrow H(v_0) - H(\sgn(v_0)) + t_0$ \;
\While{$t_1 < \theta$}{
$t_0 \leftarrow t_1$ \; 
$v_0 \leftarrow -2\sgn(v_0)$ \;
$t_1 \leftarrow H(v_0) - H(\sgn(v_0)) + t_0$ \;
}
evaluate $v_1$ such that $H(v_1) = \theta - t_0 + H(v_0)$ \tcp*{Newton-Raphson}
evaluate $v_2$ such that $f(v_2) = f(v_1)-A$\tcp*{Cubic equation (\ref{jump1})}
evaluate $v_*$, $a_1$, $a_2$, and $a_3$
\phantom{xxxxxxxxxxxxxxxxxxxxxxxxxxxxxxxxxxx}
\phantom{xxxxxxxxx}
 for the input value of $A$\tcp*{Equations (\ref{A2}) and (\ref{A3})}
$t_0 \leftarrow \theta$ \;
$v_0 \leftarrow v_2$ \;
$t_1 \leftarrow H(v_0) - H(\sgn(v_0)) + t_0$ \;
\While{$t_1 < T$}{
$t_0 \leftarrow t_1$ \; 
$v_0 \leftarrow -2\sgn (v_0)$ \;
$t_1 \leftarrow H(v_0) - H(\sgn(v_0)) + t_0$ \;
}
evaluate $v_1$ such that $H(v_1) = T - t_0 + H(v_0)$ \tcp*{Newton-Raphson}
evaluate $v_2$ such  that $f(v_2) = f(v_1)+A$\tcp*{Cubic equation (\ref{jump2})}
return $v_2$ \;
\caption{The stroboscopic map $v_{n+1} = F(v_n)$ for the singularly perturbed
FitzHugh-Nagumo system (\ref{eq}) excited by the periodic rectangular pulses (\ref{pulse})
and sampled with period $T$.}
\end{algorithm}

\end{document}